\documentclass[article,onecolumn,superscriptaddress,nofootinbib,10pt]{revtex4}
\usepackage{graphicx}
\usepackage{amssymb}
\usepackage{epstopdf}
\usepackage{amsmath}
\usepackage{hyperref}
\usepackage{mathrsfs}
\usepackage{varioref}
\usepackage{tikz}
\usepackage{lmodern}
\usetikzlibrary{decorations.pathmorphing}
\tikzset{snake it/.style={decorate, decoration=snake}}
\usepackage{fancybox}
\expandafter\ifx\csname package@font\endcsname\relax\else
\expandafter\expandafter
\expandafter\usepackage
\expandafter\expandafter
\expandafter{\csname package@font\endcsname}%
\fi
\hypersetup{
    bookmarks=false,         
    pdfstartview={FitH},    
    colorlinks=true,       
    linkcolor=red,          
    citecolor=blue,        
    filecolor=blue,      
    urlcolor=blue           
}
\DeclareFontFamily{OT1}{pzc}{}
\DeclareFontShape{OT1}{pzc}{m}{it}%
             {<-> s * [0.900] pzcmi7t}{}
\DeclareMathAlphabet{\mathscr}{OT1}{pzc}%
                                 {m}{it}
\labelformat{section}{Section #1} 
\labelformat{subsection}{Section #1} 
\labelformat{equation}{Eq.(#1)} 
\labelformat{figure}{Fig.~#1} 
\labelformat{subfigure}{Fig.~\thefigure#1} 
\labelformat{table}{Tab.~#1} 
\newcommand{\be}{\begin{equation}}
\newcommand{\ee}{\end{equation}}
\newcommand{\bea}{\begin{eqnarray}}
\newcommand{\eea}{\end{eqnarray}}

\begin{document}

\title{Unequal time Commutators in Friedmann universes : Deterministic evolution of massless fields }
\author{Kinjalk Lochan}%
\email{kinjalk@iisermohali.ac.in}%
\affiliation{Department of Physical Sciences, IISER Mohali, Manauli 140306, India}


\begin{abstract}
We analyze minimally coupled massless scalar field in a Friedmann (FRW) universe in conformal co-ordinates to model the evolution of tensor perturbations and study the structure of the Wightman function therein.  Using a duality map from a power law FRW universe to the de Sitter universe for such fields we obtain  unequal time commutation relations between quantum field variables. We demonstrate that the commutation relations are invariant under state change and/or vacuum state selection. Using such commutators it is then possible to construct out of time ordered commutators (OTOC) in the FRW universes. The OTOCs are supposed to suggest the onset of chaotic behavior during the quantum evolution, we see that  in case of Friedmann universes, unlike the scalar perturbations, the causal structure arrests the growth of quantum tensor perturbations for all the relevant epochs of the universe e.g. the de Sitter phase, the radiation dominated and the matter dominated era. Therefore the semi classical results of having a large backreaction and omnipresent noise in certain branches of the Friedmann universes with massless fields such as the tensor perturbations remain robust and stable.
\end{abstract}

\maketitle
\section{Introduction}

In the popular phrase ``{\it spacetime tells matter how to move; matter tells spacetime how to curve}'' lies the mutual impact  the matter and the gravity put on each other - one of the most prominent insights the theory of General Relativity endows us with.  Apart from the broad understanding of how curvature is to be generated with matter, in the cosmological arena the geometric perturbations are understood to give way to matter perturbations themselves, providing a concrete testimony to the aforementioned relation \cite{Brandenberger:2003vk}. 

How the quantum matter affects the spacetime and in turn how does it get affected itself is also an intriguing area of study. Analysis of quantum matter has led to various interesting insights such as particle creation, generation of primordial inhomogeneities, primordial electromagnetic and gravitational waves among many others \cite{Parker:1968mv, Parker:1969au, Parker:1971pt, Mottola:1984ar, Fulling:1989nb, Parker:1999td, Hollands:2014eia, Kobayashi:2014sga, Subramanian:2015lua}.  
Out of many observational windows available to peep into the dynamics of the early universe, the study of inflationary perturbations holds a special place. These perturbations not only provide one of the most accurate observational test-ground of the robustness of inflationary paradigm but also provide one of the very few avenues where interplay between quantum theory and gravitational dynamics  is available for ample display. In the semiclassical scheme, perturbations in the inflationary (parent) field are supposed to propagate on a background set up by the parent field itself. Being quantum mechanical in nature, these perturbations act as primordial seed of  the density fluctuations we witness in the form of galaxy clusters in the universe today. Therefore, these inflationary perturbations provide the quantum justification of the universe coming about from {\it nothing}.

The perturbations in the early universe are largely classified in terms of scalar and tensor perturbations. Vector perturbations are understood to decay very quickly to leave any impressionable imprint at recombination. The gauge-invariant scalar perturbations are effectively understood to generate the density perturbation (typically studied in terms of gauge invariant- Mukhanov Sasaki variable) whereas the gauge-invariant tensor perturbations appear as primordial gravitational waves with 2 degrees of freedom both of which effectively propagate as minimally coupled massless fields in a conformally flat setting \cite{Brandenberger:2003vk}. Therefore the individual polarization states of tensor perturbations can be modelled by massless scalar fields in Friedmann universes.

It is further conceptualized that the universe migrated from being in a quantum state to a classical state (at least statistically) \cite{Kiefer:1998qe} as the perturbations grew, thus enabling a classical analysis of the perturbations as well, forgetting about the quantum history they once had.  It is nevertheless fruitful to ask if the perturbations continue to grow will they ever be able to rekindle their quantum character and impart significant effect such as back reaction in the later part  of evolution of the universe ? In the semiclassical analysis this query is addressed by computing a quantum object called the noise kernel which is computable from the quantum variance in the stress energy tensor of the quantum matter field \cite{Hu:2008rga}. It may so happen that quantum  correlators become of as importance as the quantum expectations which play the role of classical stress energy tensor, for example, in the semiclassical analysis. It is well known that non-conformal minimally coupled massless field in de Sitter spacetime has a startling divergent correlator structure  \cite{Miao:2010vs, Krotov:2010ma, Polyakov:2012uc, Akhmedov:2012dn, Anderson:2013ila, Wang:2015eaa, Kahya:2011sy, Mora:2012kr, Mora:2012zi, Mora:2012zh, Ford:1984hs, Allen:1985ux, Antoniadis:1985pj, Allen:1987tz, Polarski:1991ek, Kirsten:1993ug, Ratra:1984yq, Dolgov:1994ra, Takook:2000qn, Tolley:2001gg, Garbrecht:2006df,  Tanaka:2013caa, Woodard:2004ut, Akhmedov:2013vka}. Though this divergent term is sometimes considered  harmless as it most likely gets removed under renormalization or derivative actions \cite{Garbrecht:2006df, Page:2012fn, Agullo:2008ka}, it is the relation between the massless fields between the power law universes to those in de Sitter space which makes the correlator structures of the fields interesting \cite{Lochan:2018pzs}. In \cite{Dhanuka:2020} it is demonstrated that conformal map between massless quantum fields in the accelerating spacetimes to quantum fields with mass in de Sitter may facilitate a strong quantum backreaction. Thus quantum correlations are expected to disrupt the classical dynamics of the background spacetime considerably and should be accounted for. This is facilitated by the long range correlation which also persists in de Sitter, but unlike in the de Sitter space they adopt a dynamic character in the Friedmann spacetimes and these correlators then become capable of generating strong second order effect in the universes of certain class  i.e., those driven by a fluid of $\omega \in[0,-1)$ \cite{Lochan:2018pzs,Dhanuka:2020}.

In this paper we analyze if the background evolution causes any significant effect on the evolution of the tensor perturbations themselves by modelling them through massless scalar fields. In the expanding universe, if the perturbations evolve in a deterministic fashion all the way, the backreaction of them on the background remains reliably robust and must be properly accounted for. However, due to large scale correlations built in them due to the omnipresent de Sitter character,  the expanding universe may lead the evolution into a chaotic regime as seen in many cases. Since in the Friedmann universe context, the long range correlations are {\it switched on} for example in the de Sitter as well as the matter dominated cases \cite{Lochan:2018pzs}, it is reasonable to check if the the large second order back reaction envisaged under deterministic evolution of the quantum field \cite{Dhanuka:2020} can be considered stable. For this purpose we analyze the correlator and the causal structure of the field commutators to analyze the late time chaotic behavior if any. In the quantum domain the ``thermalized" average (at temperature $\beta^{-1}$) of the so called out of time ordered correlator/commutators (OTOC) ${\cal C}(t,t')=Tr([O_1(t),O_2(t')]^2)_\beta$ between two Hermitian observables $O_1$ and $O_2$, are expected to carry 
the signature if any chaotic behaviour away from a quantum deterministic evolution, gets built up \cite{Maldacena:2015waa}. The study of OTOC in the cosmological scenario has been investigated in a limited fashion \cite{Haque:2020pmp,  Choudhury:2020yaa, Choudhury:2021tuu}, largely with the squeezed state analysis to obtain the signatures of chaos, and mostly via the analysis of a single mode where it is argued that the de Sitter scalar perturbations are susceptible to chaotic evolution \cite{Haque:2020pmp}.  We consider massless fields in conformal settings and employ the conformal maps between the de Sitter space and the Friedmann universes to study the possibility of chaotic evolution therein. We plan to analyze the OTOC behavior for different power law universes through a connection of the OTOCs of such universes to those in the de Sitter spacetime. Apart from illustrating some non-trivial semiclassical features such power law cosmologies are interesting from the point of view of mimicking the later epochs of our universe.  We will consider perturbations as massless fields in different epoch of evolution (i) de Sitter-inflationary phase, (ii) Radiation dominated era (iii) Matter dominated era and model each of them (crudely) with a power law evolution. Late time accelerated era again can be modelled as a de Sitter analysis.  We will see that evolution of curvature  perturbations for instance can be seen as conformally related to those in the maximal symmetric spacetimes - (i) Minkowski spacetime (ii) de Sitter spacetime. Thus the causal structures of the quantum fields in these epoch are conformally related to the causal structure in these maximally symmetric spacetimes. The conformal factors crucially mellow down the growth of OTOCs to make the semiclassical analysis and hence the statement of the large back reaction therein if any, reliable.

The paper is organized in the following fashion. In \ref{Non-Com} we review the methodology of obtaining the  relevant commutators in general spacetime through the Wightman function. We obtain the relevant commutators in the Minkowski as well as in the de Sitter spacetimes which would later be used to obtain the commutator structures in the radiation and matter dominated era. In \ref{BogoliubovWF} we demonstrate the state independence of the commutators through a Bogoliubov transformation of the vacua, which we argue will also be the case for a general Fock space state selection. In \ref{FRW-dS} we review the Wightman map connection between the power law FRW universes with the de sitter space and obtain the unequal time commutators for the radiation and matter dominated universes.  Finally, we conclude with a summary of our main findings in \ref{Summ}.


\section{The amount of non-commutativity} \label{Non-Com}
In a general spacetime, with the help of the mode functions or equivalently the Wightman function $G(x,y) $ the unequal time commutators, which we call an {\it OTOC measure}, between the field variables can be obtained  as 
\bea
[\phi(x),\phi(y)] = \int_k \int_{k'} [ \hat{a}_k u_k(x) + \hat{a}_k u_k^*(x), \hat{a}_k u_k(y) + \hat{a}_k u_k^*(y)] = \int_k\{ u_k(x) u_k^*(y)- u_k(y) u_k^*(x)\},\nonumber\\
=G(x,y)-G(y,x).
\eea
Whenever, the two points $x$ and $y$ appearing in the Wightman functions are in causal connect, the Wightman function develops an  imaginary component, which makes the field operators at any two causally communicating points non-commutative.  Since for the real scalar field, from construction we have, 
 \bea
 G(x,y)^* = G(y,x),
 \eea
 if we expand the Wightman function in its real and imaginary parts we get
 \bea
 G_R(x,y)-iG_I(x,y) = G_R(y,x) +i G_I(y,x),
 \eea
 leading to the realization that  the real part of the Wightman function has to be symmetric in $x$ and $y$, while the imaginary part must be antisymmetric :
 \bea
  G_R(x,y)&=& G_R(y,x)\nonumber\\
  G_I(x,y)&=& -G_I(y,x).
 \eea
These features are generic to Wightman function in any setting and we can obtain the commutator between the  field variables at unequal times as 
\bea
[\phi(x),\phi(y)] = G(x,y)-G(y,x) = 2i G_I(x,y) =-2i G_I(y,x).
\eea
Similarly the OTOC  measure between the conjugate variables $\phi(x)$ and $\pi(x)$ can be written as
\bea
[\phi(x),\pi(y)] =  \int_k\{ u_k(x) \dot{u}_k^*(y)- \dot{u}_k(y) u_k^*(x)\}= 2 i \partial_{y^0} G_I(x,y).
\eea
We shall now see the simple examples of such constructs in maximally symmetric spaces such as the Minkowski spacetime and the de Sitter spacetime before obtaining these in Friedmann universes of interest.


\subsection{Minkowski spacetime}
\noindent For massless scalar field $\phi$ in flat spacetime governed by the action
\bea
S=-\frac{1}{2}\int d^4 x \eta^{\alpha \beta}\left(\partial_{\alpha}\phi\partial_{\beta}\phi\right),
\eea
the mode functions as the solution to the equation of motion
\bea
\partial_{\mu}\partial^{\mu}\phi=0;
\eea
are given as
\bea
u_k(t, {\bf x}) = \frac{1}{(2\pi)^{3/2}}\frac{e^{-ikt}}{\sqrt{2k}}e^{i {\bf k} \cdot {\bf x}},
\eea
and the Wightman function is obtainable  as
\bea
G_{M}(x,y)=\frac{1}{(2 \pi)^3 }\int d^3{\bf k}\frac{e^{-ik \Delta t}}{2k} e^{i{\bf k} \cdot {\bf R}},
\eea
where $\Delta t$ and $R$ are the temporal and spatial separation between $x,y$. Performing the the ${\bf k}$- integrations we get
\bea
G_M(x,y)=\frac{-1}{4 \pi^2}\frac{1}{(\Delta t - i \epsilon)^2 -R^2} = \frac{-1}{4 \pi^2}\frac{1}{2R}\left[\frac{1}{(\Delta t -R) - i \epsilon}-\frac{1}{(\Delta t +R) - i \epsilon} \right] \nonumber\\
 = \frac{-1}{4 \pi^2}\frac{1}{2R}\left[\frac{(\Delta t -R) + i \epsilon}{(\Delta t -R)^2 +\epsilon^2}-\frac{(\Delta t +R) +i \epsilon}{(\Delta t +R)^2 + \epsilon^2} \right]  \nonumber\\= \frac{-1}{4 \pi^2}\frac{1}{(\Delta t)^2 -R^2} -\frac{ i \text{sgn}(\Delta t)}{4 \pi} \delta(\Delta t^2-R^2).\label{GreenFlat}
\eea
 This leads to a flat space commutation relation
\bea
[\phi(x),\phi(y)]= -\frac{ i \text{sgn}(\Delta t)}{2\pi} \delta(\Delta t^2-R^2) = - \frac{ i \text{sgn}(\Delta t)}{4 \pi R} \left[\delta(\Delta t-R)+ \delta(\Delta t + R)\right]. \label{CommutFlat}
\eea
The commutator between conjugate variables is, therefore,
\bea
[\phi(x),\pi(y)]= \frac{ i \text{sgn}(\Delta t)}{4 \pi R} \left[\delta'(\Delta t-R)+ \delta'(\Delta t + R)\right]= \frac{i}{2 \pi R}\left[ \frac{\delta(\Delta t -R)}{\Delta t -R} + \frac{\delta(\Delta t +R)}{\Delta t + R} \right], \label{ConjugateFlat}
\eea
which leads to the standard equal time commutation relation in the $\Delta t \rightarrow 0$ as $\delta({\bf r}) =\frac{1}{(2 \pi)^3}\int d^3{\bf k} e^{i \bf k\cdot r} =-\frac{1}{2 \pi r}\delta'(r)$.
Thus, from \ref{CommutFlat} and \ref{ConjugateFlat} we see that both the commutators are only supported {\it  on} the light cone, with no support inside or outside the light cone, for massless fields. This is natural, as the massless field modes in flat spacetime are expected to travel on null lines (as they are solution to classical equations of motions). Thus any sharing of information should occur through exchanges along the null curves. However, once we turn on the curvature of the spacetime this picture substantially changes as we shall see below for the case of de Sitter spacetime. 


\subsection{de Sitter spacetime}
\noindent For a minimally coupled massless scalar field in a Friedmann universe, 
\bea
S= -\frac{1}{2}\int d^{4}x \, a^4\big(a^{-2}\,\eta^{\alpha\beta}\partial_{\alpha}\phi \,\partial_{\beta} \phi\big)\, ,
\eea
the field has equation of motion
\bea
\ddot{\phi}+2\frac{\dot{a}}{a}\dot{\phi}+k^2\phi=0,
\eea
which is also satisfied individually by the components $(h_+,h_{\times})$ of tensor perturbations \cite{Brandenberger:2003vk}. Corresponding to this equation of motion,  in the de Sitter space $a(\eta)=-1/H\eta$, the mode-function compatible to the Bunch Davies vacuum state in the conformal time co-ordinate ($\eta$) is \cite{Parker:2009uva, Baumann:2009ds} 
\bea
v_k(\eta, {\bf x}) =\frac{1}{\sqrt{H}} \frac{(H\eta)^{3/2}}{(2\pi)^{3/2}}\frac{e^{-ik\eta}}{\sqrt{2k\eta}}\left(1-\frac{i}{k \eta}\right)e^{i {\bf k} \cdot {\bf x}}
= \frac{1}{\sqrt{2H}} \frac{1}{(2\pi)^{3/2}}\left(\frac{H}{k}\right)^{3/2}e^{-ik\eta}( k\eta -i)e^{i {\bf k} \cdot {\bf x}},
\eea
and the Wightman function is obtainable  as
\bea
G_{BD}(\eta,\eta';{\bf R})=\frac{H^2 \eta \eta'}{(2 \pi)^3 }\int d^3{\bf k}\frac{e^{-ik(\eta-\eta')}}{2k}\left[\left(1-\frac{i}{k \eta}\right)\left(1+\frac{i}{k \eta'}\right) \right]e^{i{\bf k} \cdot {\bf R}}.
\eea
Performing the the angular integrations we get
\bea
G_{BD}(\eta,\eta';{\bf R})=\frac{H^2 }{(2 \pi)^2}\int_0^{\infty} \frac{dk}{2iR} \left[\eta \eta' -\frac{i}{k}(\eta'-\eta) + \frac{1}{k^2}\right]\left(e^{-i k (\Delta \eta - R)} - e^{-i k (\Delta \eta + R)} \right),
\eea
where $\Delta \eta = \eta-\eta'$. We can split this into three independent $k-$integrations
\bea
G_{BD}(\eta,\eta';{\bf R})=\frac{H^2}{(2 \pi)^2} \left[\underbrace{\eta \eta'\int_0^{\infty} \frac{dk}{2iR}\left(e^{-i k (\Delta \eta - R)} - e^{-i k (\Delta \eta + R)} \right)}_{I_1} + \underbrace{i\Delta \eta \int_0^{\infty} \frac{dk}{2iR}\frac{1}{k}\left(e^{-i k (\Delta \eta - R)} - e^{-i k (\Delta \eta + R)} \right)}_{I_2}\right. \nonumber\\
+\left. \underbrace{\int_0^{\infty}\frac{dk}{2iR}\frac{1}{k^2}\left(e^{-i k (\Delta \eta - R)} - e^{-i k (\Delta \eta + R)} \right)}_{I_3} \right], \label{GreenFunctionBD}
\eea
and focus on the $k-$ integrations separately, which will be useful for later part of discussions as well. The first integral $I_1$ is a reminiscent of the flat space Wightman function
\bea
I_1 = \eta \eta'\int_0^{\infty} \frac{dk}{2iR}\left(e^{-i k (\Delta \eta - R)} - e^{-i k (\Delta \eta + R)} \right)=-\frac{\eta \eta'}{\Delta \eta^2 - R^2} -i \pi \eta \eta' \text{sgn}(\Delta \eta) \delta(\Delta \eta^2 - R^2). \label{I1}
\eea
The second integral is
\bea
I_2 = i\Delta \eta \int_0^{\infty} \frac{dk}{2iR}\frac{1}{k}\left(e^{-i k (\Delta \eta - R)} - e^{-i k (\Delta \eta + R)} \right) =\frac{\Delta \eta}{2R}\int_0^{\infty}\frac{d k}{k}\left(e^{-i k (\Delta \eta - R)} - e^{-i k (\Delta \eta + R)} \right). \label{I2}
\eea
We can evaluate the two terms appearing in the integrand individually. Unfortunately, both these individual terms (but the combination does not) have a pole at $k=0$. Therefore, we first put a cut-off at $k=a$ for regulating the pieces, which will be set to vanish at the end of calculations, leading to
\bea
\int_a^{\infty}\frac{d k}{k} e^{-i k (\Delta \eta \pm R)} &=& \Gamma[0,i a (\Delta \eta \pm R) ],
\eea
hence we get,
\bea
I_2(a) \equiv \frac{\Delta \eta}{2R} \int_a^{\infty}\frac{d k}{k}\left(e^{-i k (\Delta \eta - R)} - e^{-i k (\Delta \eta + R)} \right) =\frac{\Delta \eta}{2R}\left( \Gamma[0,-i a (R- \Delta \eta) ] -\Gamma[0,i a (R + \Delta \eta) ]\right), \label{I2A}
\eea
where $\Gamma[0,z]$ is an incomplete Gamma function. We will defer the dwelling  upon the properties of this function in the infrared limit $a\rightarrow 0$ till the time we calculate the third 'regularized' integral appearing in \ref{GreenFunctionBD} above
\bea
I_3(a) \equiv \int_a^{\infty}\frac{dk}{2iR}\frac{1}{k^2}\left(e^{-i k (\Delta \eta - R)} - e^{-i k (\Delta \eta + R)} \right).
\eea
This integral can simply be evaluated through integration by parts to 
\bea
I_3(a) = \frac{1}{2 i R} \left[\frac{e^{i a (R- \Delta \eta)}- e^{-i a (R + \Delta \eta)}}{a} + i (R- \Delta \eta) \Gamma[0,-i a (R- \Delta \eta) ] + i(R + \Delta \eta)\Gamma[0,i a (R + \Delta \eta) ] \right],
\eea
yielding
\bea
I_3(a) =1 + \frac{1}{2} \left(\Gamma[0,-i a (R- \Delta \eta) ]+ \Gamma[0,i a (R + \Delta \eta) ] \right)-\frac{\Delta \eta}{2 R}\left(\Gamma[0,-i a (R- \Delta \eta) ]- \Gamma[0,i a (R + \Delta \eta) ] \right) + {\cal O}(a \Delta ). 
\eea
Using the expansion of the incomplete Gamma functions
\bea
\Gamma[0,i a (R + \Delta \eta) ]&=& -\log{\left[ i a e^{\gamma} (\Delta \eta + R)\right ]} - \sum_{k=1}^{\infty} \frac{[-i a (\Delta \eta + R)]^k}{k k!}\\
\Gamma[0,-i a (R- \Delta \eta) ]&=& -\log{\left[ i a e^{\gamma} (\Delta \eta - R)\right ]}- \sum_{k=1}^{\infty} \frac{[-i a (\Delta \eta - R)]^k}{k k!},
\eea
we have
\bea
I_3(a)  = 1- \frac{1}{2} \left[\log{\left[ i a e^{\gamma} (\Delta \eta + R )\right ]}+ \log{\left[ i a e^{\gamma} (\Delta \eta - R)\right ]}\right]  -I_2(a) +{\cal O}(a \Delta), \label{I3}
\eea
with $\gamma$ being the Euler-Maclaurin constant and $\Delta$ stands collectively for the spacetime separation $\Delta\eta, R$.
Substituting the expressions for \ref{I1}, \ref{I2} and \ref{I3} in \ref{GreenFunctionBD}, we obtain the Wightman function to be
\bea
&{}& G_{BD}(\eta,\eta';{\bf R})= \frac{H^2}{(2 \pi)^2 }\left[-\frac{\eta \eta'}{\Delta \eta^2 - R^2} -i \pi \eta \eta' \text{sgn}(\Delta \eta) \delta(\Delta \eta^2 - R^2) + 1 - \frac{1}{2} \left(\log{\left[ i a e^{\gamma} (\Delta \eta + R )\right ]}+ \log{\left[ i a e^{\gamma} (\Delta \eta - R)\right ]}\right)\right]_{a \rightarrow 0},\nonumber\\
&=& \frac{H^2}{(2 \pi)^2 }\left[-\frac{\eta \eta'}{\Delta \eta^2 - R^2}   -i \pi \eta \eta' \text{sgn}(\Delta \eta) \delta(\Delta \eta^2 - R^2)+ 1 - \frac{1}{2}\log{\left[e^{i \pi \text{sgn}(\Delta \eta)\Theta(\Delta \eta^2 - R^2 )} a^2 e^{2\gamma} |(\Delta \eta^2 - R^2 )|\right]}\right]_{a \rightarrow 0} \label{GreenFunctionFinal00}.
\eea
where the cutoff scale $a$ is sent to zero as promised. As a consequence we can see that the commutator ( in the Bunch Davies vacuum state) of the field variables, at two different points $x$ and $y$ separated by $\Delta \eta^2 > R^2$ it becomes non-zero
    \bea
  [\phi(x),\phi(y)]= i \frac{ H^2\text{sgn}(\Delta \eta)}{4 \pi}\left[ \theta(\Delta \eta^2 - R^2)- 2\eta \eta' \delta(\Delta \eta^2 - R^2)\right]. \label{DSOTOC1}
    \eea
The relations \ref{GreenFunctionFinal00} and \ref{DSOTOC1} go back to their flat space counterparts \ref{GreenFlat}, \ref{CommutFlat} for $H\rightarrow 0$ as the relation between the conformal and comoving times $\eta^{-1} = -He^{Ht}$ ensures the dropping of the curvature induced terms. 
Thus we see that for massless fields, the commutator structure $[\phi(x),\phi(y)]$ obtains a curvature supported uniform value inside the light cone. Clearly this value does not depend on the separation $\eta-\eta'$ and hence the initial commutation, once set inside the light cone does not decay. Thought this measure does not grow in time either, therefore is not capable of setting up a quantum chaos.
We can also evaluate the OTOC measure between the conjugate variables $\phi$ and $\pi =a^2 \dot{\phi}$ (for a scalar field in Friedmann space in conformal co-ordinates)
 \bea
  [\phi(x),\pi(y)] &=&\int d^3{\bf k} a^2(\eta')[ u_k (x)\dot{u}_k^{*}(y) - u_k^* (x)\dot{u}_k(y) ].
  \eea
Since 
\bea
\dot{u}_k(x) = \frac{1}{\sqrt{2H}} \frac{1}{(2\pi)^{3/2}}\left(\frac{H}{k}\right)^{3/2}e^{-ik\eta}(-i k\eta^2)e^{i {\bf k} \cdot {\bf x}},
\eea
using the expressions for \ref{I1}, \ref{I2A} and \ref{I3} we can write
   \bea
  [\phi(x),\pi(y)] &=&  \frac{H\eta}{H\eta'}\frac{i}{2\pi R}\left[\frac{\delta(\Delta \eta - R)}{\Delta \eta - R} - \frac{\delta(\Delta \eta + R)}{\Delta \eta + R}\right] -i\frac{H}{2 \pi H\eta'}\text{sgn}(\Delta \eta)\delta( \Delta \eta^2 - R^2).
\label{DSOTOC2}
    \eea
 The field field OTOC measure was uniformly supported inside the light cone whereas the conjugate variables has a support only on the light cone. Yet, none of these measures grow. Thus, the tensor perturbations behave somewhat different compared to the gauge invariant scalar perturbations \cite{Haque:2020pmp}.
Unlike the scalar perturbations the tensor perturbations in de Sitter spacetime do not grow to develop any chaotic behavior, and infact decay with the growth of the scale factor $a\sim 1/\eta$. The de Sitter spacetime already suffers with uncomfortable characteristics for massless fields, e.g. the divergent correlation, secular growth or large potential backreaction \cite{Miao:2010vs, Krotov:2010ma, Polyakov:2012uc, Akhmedov:2012dn, Anderson:2013ila, Wang:2015eaa, Kahya:2011sy, Mora:2012kr, Mora:2012zi, Mora:2012zh, Ford:1984hs, Allen:1985ux, Antoniadis:1985pj, Allen:1987tz, Polarski:1991ek, Kirsten:1993ug, Ratra:1984yq, Dolgov:1994ra, Takook:2000qn, Tolley:2001gg, Garbrecht:2006df,  Tanaka:2013caa, Woodard:2004ut, Akhmedov:2013vka}, the analysis of the OTOC measures for conformal massless fields such as the tensor perturbations keeps the semiclassical analysis steady.

As discussed above, the Friedmann universes with power law evolution share many similarities with the de Sitter spacetime and in fact for conformal massless fields the dynamics in a Friedmann universe can exactly be mapped to that of a massive field in the de Sitter spacetime, it will be worthwhile to check whether or not, the growth of OTOC measures is capable of potentially disrupting the semiclassical analysis in such spacetimes. 

Before we proceed to evaluate such measures for the other FRW phases of the universe it is helpful that we establish the state independence of such measures. We have evaluated these objects for correlator in Bunch Davies vacuum but it is straight forward to demonstrate that the commutator structure is really independent of the choice of vacuum or even choice of state in the chosen Fock basis, which we do next. Thus the usual prescriptions of removal of zero modes, infrared regularization of the theory \cite{Garbrecht:2006df, Page:2012fn, Agullo:2008ka} to arrest the secular growth of the field still does not cure the 'susceptibility' of the de Sitter space against onset of quantum chaos.

\section{Wightman function : Change of states} \label{BogoliubovWF}

A Wightman function is a generally covariant entity, therefore it will maintain its structure once derived in any frame. Still if we wish to connect Wightman function computed in one observer's frame to another set of observers related through non trivial Bogoliubov coefficients, we need to transform one into another. In the vacuum associated with the mode functions $u_k(x)$, the Wightman function is
\bea
G_u(x,y)\equiv \langle 0 | \hat{\phi}(x) \hat{\phi}(y) | 0 \rangle  = \sum_k u_k(x) u_k^*(y).
\eea
If we use mode function $v_k(x)$ related to $u_k(x)$ through
\bea
v_k(x) &=& \sum_{k'}(\alpha_{k k'}u_{k'}(x) + \beta_{k k'}u^*_{k'}(y)),
\eea
with the Bogoliubov coefficients $\alpha_{k k'}$ and $\beta_{k k'}$ satisfying the following identities
\bea
\int dk (\alpha_{k' k}\alpha^*_{k^{''} k}- \beta_{k' k}\beta^*_{k^{''} k}) &=& \delta(k'-k^{''}) \label{BogoCoff1} \\
\int dk (\alpha_{k' k}\beta_{k^{''} k}- \beta_{k' k}\alpha_{k^{''} k}) &=& 0.\label{BogoCoff2}
\eea
If the new mode functions are  on (even a portion of) a Cauchy surface, the following is also true
\bea
u_k(x) = \int dk' ( \alpha^*_{k' k}v_{k'}(x) -\beta_{k' k}v^*_{k'}(y)),
\eea
however, leading to identities which will be obeyed only if one has the new set of modes defined on a Cauchy surface
\bea
\int dk (\alpha_{k k'}\alpha^*_{ k k^{''}}- \beta_{ k k'}\beta^*_{ k k^{''}}) &=& \delta(k'-k^{''}) \\
\int dk (\alpha^*_{ k k'}\beta_{ k k^{''}}- \beta_{ k k'}\alpha^*_{ k k^{''}}) &=& 0. \label{InvBogoCoff}
\eea
Therefore,
\bea
G_u(x,y) =
\int dk dk' dk^{''} \left[\alpha^*_{k' k}  \alpha_{k^{''} k}v_{k'}(x)v^*_{k^{''}}(y)-\beta_{k' k} \alpha_{k^{''} k}v^*_{k'}(x)v^*_{k^{''}}(y)\right.\nonumber\\
\left.-\alpha^*_{k' k}  \beta^*_{k^{''} k}v_{k'}(x)v_{k^{''}}(y)+\beta_{k' k}  \beta^*_{k^{''} k}v^*_{k'}(x)v_{k^{''}}(y)\right].
\eea
Using the relations \ref{BogoCoff1}, \ref{BogoCoff2} we can write the above expression as
\bea
G_u(x,y) &=& \int dk' v_{k'}(x)v^*_{k^{'}}(y) + \int dk \left[\left(\int  dk^{''} \beta_{k^{''} k}v^*_{k^{''}}(y)\right)  \left(\int  dk^{'} \beta^*_{k^{'} k}v_{k^{'}}(x)\right)  \right. \nonumber\\
&+& \left. \left(\int  dk^{''} \beta^*_{k^{''} k}v_{k^{''}}(y)\right)  \left(\int  dk^{'} \beta_{k^{'} k}v^*_{k^{'}}(x)\right)\right]
- \int dk \left[\left(\int  dk^{''} \alpha_{k^{''} k}v^*_{k^{''}}(y)\right)  \left(\int  dk^{'} \beta_{k^{'} k}v^*_{k^{'}}(x)\right) \right.\nonumber\\
&+& \left. \left(\int  dk^{''} \alpha^*_{k^{''} k}v_{k^{''}}(y)\right)  \left(\int  dk^{'} \beta^*_{k^{'} k}v_{k^{'}}(x)\right)\right]
\eea
The first term in the RHS is the Wightman function corresponding to the vacuum of $v_k(x)$. In addition there are correction terms which are  real in total (therefore the commutator remains protected in Bogoliubov transformation). These extra terms can be neatly combined in the form
\bea
G_u(x,y) &=& \int dk' v_{k'}(x)v^*_{k^{'}}(y) -\left [ \int dk u_k(y) \int dk' \beta^*_{k'k}v_{k'}(x) + c.c \right].
\eea
If we define a new quantity made up of superposition of the positive frequency mode functions of the new basis
\bea
\tilde{v}_{k}(x) =\int dk \beta^*_{kk'}u_{k'}(x), \label{NuU}
\eea
one can verify  that
$$ (\tilde{v}_{k'}, \tilde{v}_{k'}) = \int dk|\beta_{k'k}|^2 = N_{k'},$$
where $N_k$ is the occupancy of the mode $k$ in the mode $v_k$.

Using \ref{NuU} we have
\bea
G_u(x,y) &=& \int dk' v_{k'}(x)v^*_{k^{'}}(y) -2\text{Re}\left[\int dk' v_{k'}(x)\tilde{v}_{k'}(y) \right],
\eea
which can be written as
\bea
G_u(x,y) &=& G_v(x,y) -2\text{Re}\left[\int dk' v_{k'}(x)\tilde{v}_{k'}(y) \right]. \label{WightmanDiff}
\eea
One can clearly see that the two Wightman functions can differ at most by a real quantity. Therefore, the commutator structure will not be different for these two vacua.  The structure can similarly be generalized for non-vacuous state of the field \cite{Lochan:2014xja}. 
In fact, for a massive scalar field, one can obtain a de Sitter invariant vacuum in which the Wightman function assumes the form \cite{Page:2012fn}
\bea
G_{m, dS}(x,y) = \frac{H^2 \Gamma [h_+]\Gamma [h_-]}{16\pi^2}{}_2F_1(h_+,h_-,2,1-z)
\eea
where $h_{\pm}= 3/2\pm\sqrt{9/4-m^2}$ and ${}_2F_1(a,b,c,x)$ is the hyper-geometric function. Typically well-behaved Wightman functions have a pole at the light cone (can be seen from its Hadamard form) which contributes to an imaginary (non-symmetric contribution) to the Wightman function. For a scalar field (with $m^2\rightarrow 0$) in the Bunch Davies vacuum, we have
\bea
G(x,y) = \frac{H^2}{16 \pi^2}\frac{6}{ m^2}\big|_{m^2\rightarrow 0} +\frac{H^2}{16 \pi^2}\left( \frac{1}{z(x,y)}-2\log{z(x,y)} \right) +{\cal O}(m^2),
\eea
where $z(x,y) =\eta_{ab}[X^a(x)-X^a(y)][X^b(x)-X^b(y)]/4$ being the invariant distance between two points $(x,y)$ on a de Sitter hyperboloid \cite{Allen:1987tz}. Again, for timelike related events $z(x,y)<0$ and we have
\bea
\log{z} \rightarrow \log{|z|} +i \text{sgn}(\eta-\eta')\pi,
\eea
leading to an imaginary part in the Wightman function which subsequently leads to a non-vanishing commutation relation. Thus we can evaluate the commutator structure for any Wightman function eventually leading to the  same commutator structure as in \ref{DSOTOC1}. Now we can use the de Sitter and Minkowski commutator structures to obtain OTOC measures for other epochs of the universe.


\section{FRW Universe} \label{FRW-dS}
\subsection{Conformal map to de Sitter}

The action of a minimally coupled  massless\footnote{This analysis can be extend to any non-minimally but non-conformally coupled fields as well \cite{Lochan:2018pzs}.} scalar field in a Friedmann universe with metric, $g_{\alpha\beta}= a^2(\eta) \eta_{\alpha\beta}$ with $a(\eta) = (H\eta)^{-q}$, is given by
\begin{equation*}
S= -\frac{1}{2}\int d^{4}x \, a^4\big(a^{-2}\,\eta^{\alpha\beta}\partial_{\alpha}\phi \,\partial_{\beta} \phi\big)\, .
\end{equation*}
Under the conformal transformation, $\phi(x) = (H\eta)^{-1+q}\, \psi(x)$, the action can be written as that of a massive scalar field in another Friedmann universe with scale factor $b(\eta) = (H\eta)^{-1}$,
\begin{equation*}
S= -\frac{1}{2}\int d^{4}x \,b^4\big(b^{-2}\,\eta^{\alpha\beta}\partial_{\alpha}\psi \,\partial_{\beta} \psi - m_{eff}^2 \psi^2 \big)\, ,
\end{equation*}
where $m^2_{eff} = H^2(1 - q)(2 + q)$. Therefore, we see that a massless scalar field $\phi_{FRW}(x)$ in a Friedmannn universe with scaling factor, $a(\eta) = (H\eta)^{-q}$, goes to a massive scalar field $\phi_{dS}(x)$ with mass $m_{eff}$ in a de Sitter universe under such a conformal transformation \cite{Lochan:2018pzs}. For the de Sitter case, obviously $q=1$ whereas for radiation and the matter dominated era $q=-1,$ and $q=-2$ respectively. This leads to a description of the massless field in these two spacetimes in terms of massive fields in de Sitter spacetime with $\nu\equiv \sqrt{9/4-m_{eff}^2/H^2} =1/2$ and $\nu=3/2$ respectively. The case for the radiation dominated era (corresponding to $\nu=1/2$) is somwhat straight forward as its equivalent de Sitter description is conformally flat. We will utilize this property to analyze the correlator and the commutator structure in the radiation dominated era relating them to those in flat spacetime.

\subsection{Commutator Map}

 We also see that the transformation relation between the fields i.e., $\phi_{FRW}(x) = (H\eta)^{-1+q}\, \phi_{dS}(x)$ explains the relation between the Wightman functions in the related space-times i.e., $G_{FRW}(x,y) = (H\eta)^{q-1}(H\eta')^{q-1}G_{dS}(x,y)$. A similar kind of correspondence can be established for non-minimal coupling to gravity as well (see Appendix A.2 of \cite{Lochan:2018pzs} for details).
 
Using the duality between the massless field in the Friedmann universe $a(\eta)= (H \eta)^{-q}$ where $H$ is a constant parameter,
\bea
G_{m=0, FRW}(x,y)  = (H^2\eta \eta')^{q-1}G_{m^2=H^2(1-q)(2+q), dS}(x,y) 
\eea
we get
\bea
[\phi_{FRW}(x),\phi_{FRW}(y)] = (H^2\eta \eta')^{q-1}[\phi_{dS}(x),\phi_{dS}(y)]_{m^2=H^2(1-q)(2+q)}.
\eea
Moreover, since $\pi_{\text{FRW}}  =a^2 \dot{\phi}_{\text{FRW}}$ and $\phi_{\text{FRW}} =  \left(H \eta\right)^{q-1} \phi_{dS}$, one can also obtain the OTOC measure between the conjugate variables 

\bea
 [\phi_{\text{FRW}}(x),\pi_{\text{FRW}}(y)] =\left(H \eta \right)^{q-1}\left[(q-1)H\left(H \eta'\right)^{-q-2}[\phi_{dS}(x),\phi_{dS}(y)] +\left(H \eta'\right)^{1-q}[\phi_{dS}(x),\pi_{dS}(y)]\right]_{m^2=H^2(1-q)(2+q)}
 \eea

We can now analyze the different cosmological era separately by judiciously selecting the parameter $q$, which will fix the mass of the de Sitter field in the equivalent description.


\subsubsection{Radiation dominated era}
The radiation dominated era of the universe is identified as the scale factor with $q=-1$
leading to  $m^2/H^2=2$ or equivalently $\nu=1/2$ which is the conformal mass in de Sitter since correlator in this configuration is conformally related to that of a massless field in the flat space time, i.e.,
\bea
G_{Radiation}\left(\eta,\eta'; {\bf R}\right) = (H^2\eta \eta')^{-2}G_{dS}\left(\eta,\eta'; {\bf R};\nu=\frac{1}{2}\right);
\eea
and
\bea
G_{dS}\left(\eta,\eta'; {\bf R};\nu=\frac{1}{2}\right) = (H^2\eta \eta')G_{M}(\eta,\eta'; {\bf R};m=0).
\eea
Therefore, the correlator and the commutator structure of massless field in the radiation dominated era of the universe is just given as the conformal scaling of their Minkowskian analogue

\bea
 [\phi_{\text{Radiation}}(x),\phi_{\text{Radiation}}(y)]=(H^2\eta \eta')^{-1} [\phi_{M}(x),\phi_{M}(y)]
\eea
and
\bea
[\phi_{\text{Radiation}}(x),\pi_{\text{Radiation}}(y)]=\eta^{-1}\left( -[\phi_{M}(x),\phi_{M}(y)] +\eta' [\phi_{M}(x),\pi_{M}(y)] \right).
\eea

Thus we see that both the OTOC measures are given in terms of flat space measures and have support only on the light cone. Despite the spacetime having a non-zero curvature the causal structure of a massless field is similar to that in the flat spacetime. This is understandable as the Ricci scalar is vanishing for the radiation dominated era and the deviation of Wightman function from that of its flat space avatar is codified in terms of expansion in Ricci scalar for a Hadamard state \cite{Parker:2009uva}. Therefore, inside the light cone the evolution remains constrained only to null connected points, and the OTOC measure decays in $\eta$, i.e. becomes less correlated with growing universe $a\sim \eta$. This decay is effectively caused by the conformal factor conformal factor $(H \eta)^{q-1}|_{q=-1} $ which connects the radiation dominated and the de Sitter universes. Thus massless field's quantum treatment remains stable and trustworthy for a radiation dominated universe.


\subsubsection{Matter dominated era}
The  case of matter dominated universe is of particular interest as this spacetime shares a remarkable resemblance to the de Sitter spacetime. Owing to the relation $m^2=H^2(1-q)(2+q)$ we can see that a massless field in the matter dominated universe $(q=-2)$ maps to massless field in the de Sitter spacetimes as well. Thus, the matter dominated universe shares the very similar correlator pathologies for massless field  which the de Sitter suffers with.
In fact, it can be argued that due to the conformal connection associating a dynamics (i.e. spacetime dependence) to them, the divergences appearing in matter dominated universes are far more difficult to tackle, leading to much severe and serious quantum backreaction, unlike the case of the de Sitter \cite{Dhanuka:2020}. It is therefore, instructive to check if such an analysis of large back reaction  is really stable under potential onset of quantum chaos.

The two  OTOC measures of the de Sitter spacetime contribute to the OTOC measure of the matter dominated universe leading to 
 \bea
 [\phi_{\text{Matter}}(x),\phi_{\text{Matter}}(y)] &=& i\left(H \eta \right)^{q-1}\left(H \eta' \right)^{q-1}\frac{ H^2\text{sgn}(\Delta \eta)}{4 \pi}\left[ \theta(\Delta \eta^2 - R^2)- 2\eta \eta' \delta(\Delta \eta^2 - R^2)\right]_{q=-2},
\eea
and 
\bea
  [\phi_{\text{Matter}}(x),\pi_{\text{Matter}}(y)] &=&  (H \eta)^{q-1}\left[ (q-1) i H(H\eta')^{-q-2}\frac{ H^2\text{sgn}(\Delta \eta)}{4 \pi}\left( \theta(\Delta \eta^2 - R^2)- 2\eta \eta' \delta(\Delta \eta^2 - R^2)\right) \right. \nonumber\\
   &+& \left.(H\eta')^{1-q}\left(\frac{H\eta}{H\eta'}\frac{i}{2\pi R}\left[\frac{\delta(\Delta \eta - R)}{\Delta \eta - R} - \frac{\delta(\Delta \eta + R)}{\Delta \eta + R}\right] -i\frac{H}{2 \pi H\eta'}\text{sgn}(\Delta \eta)\delta( \Delta \eta^2 - R^2)\right)\right]_{q=-2}.
    \eea
Due to the fact that the de Sitter field field commutator has a support inside the light cone, the matter dominated universe commutator also develops a support in the interior of  the light cone
\bea
 [\phi_{\text{Matter}}(x),\phi_{\text{Matter}}(y)]_{\text{interior}} &=& i\left(H \eta \right)^{-3}\left(H \eta' \right)^{-3}\frac{ H^2\text{sgn}(\Delta \eta)}{4 \pi} \theta(\Delta \eta^2 - R^2),
 \eea
 and
 \bea
 [\phi_{\text{Matter}}(x),\pi_{\text{Matter}}(y)]_{\text{interior}} &=&-\frac{3iH^3}{4\pi}\text{sgn}(\Delta \eta)(H\eta)^{-3}\theta(\Delta \eta^2 - R^2). \label{MatterCommutator}
 \eea

Since the massless field in the matter dominated era is related to massless field in  the de Sitter space, their OTOC measures' structures are also somewhat similar. Foremost,  like the de Sitter space, the OTOC measures develop within light cone  support. Further, the Eq. \ref{MatterCommutator} shows that within the light cone, for a future point $\eta,$ the effect for conjugate variables, is uniform for all points $\eta'$ of the past, lying inside the light cone, however, with growth in the scale factor ($a \sim \eta^2$) both the OTOC measures decay as $\eta^{-3}$ for $\eta \rightarrow \infty$. This decay is basically sourced by the conformal factor $(H \eta)^{q-1} $ which effectively connects the matter dominated universe to the de Sitter space. Thus the chaos is not set up with any efficiency and the conclusions based upon the deterministic quantum treatment remains valid. As previously discussed, a stable deterministic quantum evolution of scalar massless perturbations is expected to lead to significant backreaction \cite{Dhanuka:2020} due to its divergent infrared correlation structure shared with the de Sitter. Thus the analysis suggests the robustness of large backreaction in the matter dominated era, at least from the point of view of quantum determinism. 

\section{Conclusion} \label{Summ}
The analysis of quantum fields on the expanding background leads to many profound results ranging from particle creation to realization of quantum fluctuations into matter perturbations. The analysis of massless field in such expanding backgrounds are of more interest  in the spirit that the tensor perturbations satisfying the same equation of motion, generate primordial inhomogeneity and primordial gravitational waves \cite{Brandenberger:2003vk}. Therefore the role of their quantum characteristics become  natural queries in the semiclassical analysis.
The de Sitter space suffers with pathological features if the massless fields are put on it. This leads to various discomforting characteristics. Furthermore, the massless fields in other Friedmann universes of certain class are also tightly coupled to quantum fields in the de Sitter space and many of these share such pathological features, in a much severe form \cite{Lochan:2018pzs, Dhanuka:2020}.

In this paper, we analyzed the stability of the semiclassical analysis in Friedmann universes against the onset of potential quantum chaos by evaluating the out of time ordered commutators for field variables. For the de Sitter spacetime it was demonstrated that while the uniform curvature sets up a uniform within light cone commutator structure, the tensor perturbations do not develop any chaotic tendency as reported for scalar perturbations \cite{Haque:2020pmp}. Thus the semiclassical pathologies associated with such fields in the de Sitter space, such as secular growth \cite{Ford:1984hs} and large back reaction \cite{Dhanuka:2020} remain reliable and hence of concern.

Further, the conformal map between the massless fields in power law universes to the fields in de Sitter suggests that the semiclassical analysis therein also are reliable as far as the tensor perturbations are concerned. Radiation dominated universe shares its quantum character with that of a flat spacetime while the matter dominated universe relates to a de Sitter spacetime. The dilution of the conformal factor with the growth of the universe successfully arrests any growth of potential quantum chaos.  Therefore, the semiclassical features in these spacetimes are more stable and the issue of large correlations and back reaction may need a further careful inspection.  Though in this study, we have focussed on the universes mimicking the different epochs of standard cosmology, it is further instructive to check  if any of the spacetimes with a large noise kernel $\omega \in[0,-1/3)\cup(-1/3,-1)$ \cite{Dhanuka:2020} do develop a chaotic character at late times. Furthermore in this discussion we have focussed on Friedmann universes with a  fixed equation of state (and hence with definite power law index $q$) , i.e., dominated by a single fluid. How the OTOC measures behave in a universe with multiple components is an interesting question. Furthermore, the analysis presented here avoids any self coupling of the quantum field. Such studies for self interacting theories \cite{Stanford:2015owe} can be developed from taking the duality map analysis to the interacting case \cite{Lochan:2018pzs}. However such discussions will be taken up in a subsequent study.


\section{Acknowledgements}
Research of K.L. is partially supported by the Startup Research Grant of SERB, Government of India (SRG/2019/002202). The author fondly remembers, many engaging discussions with Prof. Padmanabhan on various aspects of quantum fields in cosmology and related issues. 

\section{Data Availability}
Data sharing not applicable to this article as no datasets were generated or analysed during the current study.



\end{document}